\documentclass[prb,twocolumn,showpacs,preprintnumbers,amsmath,amssymb]{revtex4}
\usepackage{graphicx}
\usepackage{dcolumn}
\usepackage{bm}
\usepackage[latin1]{inputenc}
\usepackage[T1]{fontenc}
\usepackage[francais]{babel}
\usepackage{comment}
\usepackage{epstopdf}
\begin{document}

\title{Photophysics of single nitrogen-vacancy centers in diamond nanocrystals}
\author{Martin Berthel$^{1,2}$, Oriane Mollet$^{1,2,3}$, G\'eraldine Dantelle$^{1,2,4}$, Thierry Gacoin$^4$, Serge Huant$^{1,2}$ and Aur\'elien Drezet$^{1,2}$ }
\affiliation{$^{1}$ Universit\'e Grenoble Alpes, Inst. NEEL, F-38000 Grenoble, France\\
$^{2}$ CNRS, Inst. NEEL, F-38042 Grenoble, France\\
$^{3}$ LPN-CNRS, Route de Nosay 91460 Marcoussis, France\\
$^{4}$ Laboratoire de Physique de la Mati\`ere Condens\'ee, Ecole Polytechnique, UMR CNRS 7643, 91128 Palaiseau, France}

\begin{abstract}
A study of the photophysical properties of nitrogen-vacancy (NV) color centers in diamond nanocrystals of size of 50~nm or below is carried out by means of second-order time-intensity photon correlation and cross-correlation measurements as a function of the excitation power for both pure charge states, neutral and negatively charged, as well as for the photochromic state, where the center switches between both states at any power. A dedicated three-level model implying a shelving level is developed to extract the relevant photophysical parameters coupling all three levels. Our analysis confirms the very existence of the shelving level for the neutral NV center. It is found that it plays a negligible role on the photophysics of this center, whereas it is responsible for an increasing photon bunching behavior of the negative NV center with increasing power. From the photophysical parameters, we infer a quantum efficiency for both centers, showing that it remains close to unity for the neutral center over the entire power range, whereas it drops with increasing power from near unity to approximately 0.5 for the negative center. The photophysics of the photochromic center reveals a rich phenomenology that is to a large extent dominated by that of the negative state, in agreement with the excess charge release of the negative center being much slower than the photon emission process.
\end{abstract}
 \maketitle
\section{Introduction}
With the development of photonic quantum cryptography and quantum
information processes, there is need for reliable and easy-to-use
single photon sources. Such sources have been developed in
recent years~\cite{singlephoton}, like single molecules, colloidal or epitaxial
semiconductor quantum dots, and color-centers in diamond.~\cite{Gruber97,Prawer,Si-N} Here, we
are interested in the latter, namely the NV center, formed by a substitutional nitrogen atom adjacent to a
vacancy in the diamond lattice. NV centers have found numerous applications recently thanks to their unique physical
properties such as excellent photostability~\cite{Gruber97,Brouri00,Sonnefraud08,Bradac2010} and long spin
coherence times~\cite{Balasubramanian09} as well as to improved control
over both their production~\cite{Rondin10} and physical
initialization protocols.~\cite{Siyushev13} NV centers can be made
available both in ultra-pure bulk diamond~\cite{Balasubramanian09}
and ultra-small crystals.~\cite{Smith08} Applications range from
high-sensitivity high-resolution
magnetometry~\cite{Degen08,Maze08,Balasubramanian08,Maletinsky12,Rondin12,Rondin14,Grinolds13}, to fluorescence probing of
biological processes ~\cite{Faklaris09,McGuinness11}, solid-state
quantum information processing~\cite{Schuster10,Kubo10}, spin
optomechanics~\cite{Arcizet11,Hong12}, quantum
optics~\cite{Beveratos01,Kurtsiefer00,Sipahigil12},
nanophotonics~\cite{Schietinger09,Cuche09,Beams13,Schell}, and quantum
plasmonics.~\cite{Kolesov09,Cuche10,Mollet12}\\
NV centers can take two different charge states with different
spectral properties: the neutral center NV\up{0}, which has a
zero-phonon line (ZPL) around 575~nm (2.16 eV), and the negatively
charged center NV\up{-}, which has a ZPL around 637~nm (1.95
eV).~\cite{Dumeige04} In addition to ZPLs, the fluorescence spectra
of both centers exhibit a broad and intense vibronic band at lower
energy. A single NV center, as a single-photon emitter, is characterized by a second-order time-intensity correlation function that exhibits a photon antibunching dip
at zero delay. In a first approach, the photophysics of NV centers can be modeled by a two-level system with two
photophysical parameters, the excitation rate, and the spontaneous
emission rate. However, with increasing excitation power, the NV
center, more particularly in the negative state NV\up{-}, can
experience distinctive photon bunching at finite coincidence time,
in addition to the expected antibunching at zero delay. This can be
accounted for within a three-level system with additional
photophysical parameters to describe photon decays to,
or from, the additional shelving level.\\
The aim of this paper is to give a detailed description of the intrinsic photophysics of
single NV centers of both charge states in surface-purified~\cite{Rondin10}  nanodiamonds (NDs) of size
around 50~nm, or below, as a function of the illumination power from a CW laser. Understanding the intrinsic photophysics of NV centers is required before implementing small fluorescent NDs in a complex electromagnetic environment, such as practical single-photon
devices, which will modify the photophysics.~\cite{Schietinger09} It is also useful to the applications mentioned above as most of them exploit the single-photon emitter nature of isolated NVs.\\
The statistics of both NV charge states has been studied previously in ND samples similar to those studied here and it was found to be size dependent, with a larger occurrence (> 80\%) of the NV\up{-} center over the entire size range from 20 to 80~nm.~\cite{Rondin10}  In the present study, we further observed that most of the single NV centers being detected as neutral from their fluorescence spectrum at low illumination power (<0.5~mW) progressively gain with increasing power a photochromic character in that they also exhibit the NV\up{-} ZPL in addition to the NV\up{0} ZPL. Only a few of NV\up{0}s remain purely neutral over the entire power range, which we call pure NV\up{0} behavior. On the other hand, NV centers detected as negatively charged at low power remain so with increasing power, which we call pure NV\up{-} behavior. We first focus our attention on such pure NV\up{0} or NV\up{-} centers. In addition to such behaviors, we found that some rare NVs can see their charge switching between neutral and negative~\cite{Gaebel06} already at the lowest excitation power. We also describe the photophysics of such a photochromic NV over the same power range as the non-photochromic centers and show how both photophysics can be linked. This allows us to find valuable information on the dynamics of photochromism.\\
The paper is organized as follows. The experimental methods are described in Section II. The three-level model used to interpret the experiment is developed in detail in Section III.
Section IV focuses on the experimental results and on the extraction of the various photophysical parameters as a function of the excitation power for a NV\up{0} and a NV\up{-} center. Section V describes the photophysics of photochromism in a single NV. A summary is given in Section VI.
\section{Experimental methods}
\begin{figure}[hbtp]
\includegraphics[width=0.7\columnwidth]{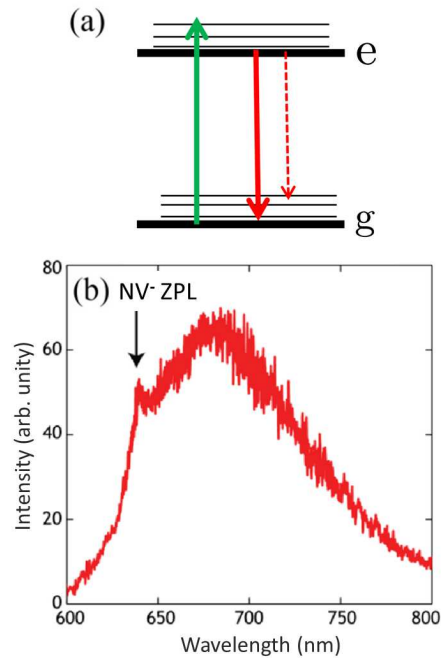}\\
\caption{ (Color online) (a) Schematic of the two-level system explaining the
emission spectrum of the NV center. g : ground state, e : excited
state. The several thinner horizontal lines represent the coupling
with the phonon bath in the diamond matrix. The green arrow
represents the excitation and the red ones the optical decay at
resonance (full arrow) and out of resonance (dotted arrow) (b)
Typical emission spectrum of an NV\up{-} center in a 25~nm
ND.} \label{Spectre}
\end{figure}

\begin{figure}[hbtp]
\includegraphics[width=0.9\columnwidth]{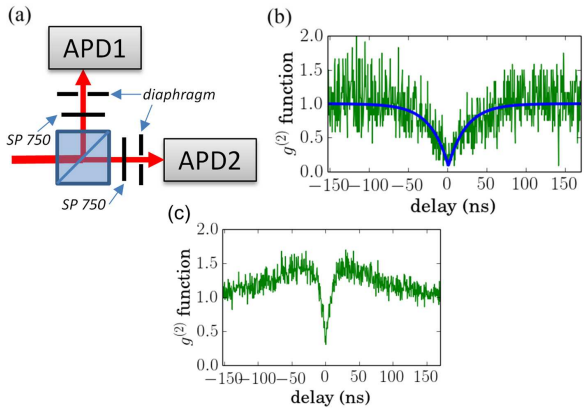}\\
\caption{ (Color online) (a) Schematic of the HBT correlator (b)
Time-intensity second-order correlation function $g^{(2)}(\tau)$ of
a NV\up{0} center excited at 200 $\mu$~W and presenting an
antibunching at zero delay,  (c) Time-intensity second-order
correlation function $g^{(2)}(\tau)$ of a NV\up{-} center excited at
5~mW  and presenting a bunching at finite delay together with the antibunching at zero delay. SP 750 : short-pass filter at 750~nm} \label{HBT}
\end{figure}

Preparation of the ND sample was achieved following a procedure
reported previously.~\cite{Dantelle10,Rondin10} Commercial HPHT
diamond nanocrystals are first irradiated using high-energy
electrons, then annealed at 800 \degres C in vacuum to produce the
fluorescent NV centers, and finally annealed in air at 550 \degres C
to remove surface graphitic compounds. Colloidal dispersion in water
and further sonication allow us to obtain a uniform solution of NDs.
We consider here NDs with a typical size of 25~nm or 50~nm deposited
on a fused silica substrate to minimize spurious
fluorescence.~\cite{Cuche09} Single NV centers were optically
addressed using standard confocal microscopy at room
temperature.~\cite{Cuche09} A CW laser light (wavelength $
\lambda_{exc} $ = 515~nm) falling within the absorption band of the
NVs is used to excite the NV fluorescence. Excitation light was
focused onto the sample using an oil immersion microscope objective
with numerical aperture NA= 1.4. The NV fluorescence is collected
through the same objective and is filtered from the remaining
excitation, i.e., with wavelength below 532~nm, by a dichroic mirror
and a high-pass filter. The collected fluorescence is subsequently
sent either to a Hanbury-Brown and Twiss (HBT) intensity correlator
(see below) or to a spectrometer. An example of fluorescence
spectrum is shown in Fig. \ref{Spectre}(b) for the NV\up{-} case.
The ZPL corresponds to the resonant decay (Fig. \ref{Spectre}(a)) at
$\lambda=637$~nm while
the wide fluorescence side-band is the phonon replica.\\
The second-order time-intensity correlation function contains the information on the classical {\it versus} quantum nature of light. It reads in the
stationary regime \begin{equation} g^{(2)}
(\tau)=P_{2}(t+\tau|t)/P_1(t)\end{equation} where
$P_2(t+\tau|t)=P_2(\tau|0)$ is the conditional probability to detect
a photon at time $t+\tau$ knowing that another photon has been
recorded at time $t$. This probability is normalized by the constant
single photon detection rate $P_1(t)=P_1(0)$. For a classical source
of light $g^{(2)} (\tau)\geq 1$, whereas the
observation of an anti-bunching $g^{(2)} (\tau)\leq 1$ is a clear
signature of the quantum nature of light.~\cite{Loudon,Mandel95,Kimble} In particular, at zero delay $g^{(2)} (0)=0$ for a single photon emitter~\cite{Kimble}, which means that the probability to detect simultaneously two photons vanishes.\\
In practice, an HBT correlator (Fig.~\ref{HBT}(a))~\cite{Brouri00,Sonnefraud08} is used to measure $g^{(2)}$. Here, the fluorescence of the NV center is sent on a beam splitter, which separates the signal in two equal parts sent to two avalanche photo-diodes (APDs) named APD1 and APD2. The APDs are connected to a time-correlated single-photon counting module to build histograms of delays between photon events detected by the upper \og Start \fg{} APD1 and the lower \og Stop \fg{} APD2. In order to avoid unwanted optical crosstalk between the APDs, a glass filter acting as a short-pass filter at 750~nm and a diaphragm are added in both branches.~\cite{Kurtsiefer01} In the standard configuration, no bandpass filter is added to the setup, in contrast with the configurations used to study photochromism (Section V), so that it can be used for both charge states of the NV (provided that the NVs do not experience charge conversion, which is the case for the selected NVs in the present work).\\
\section{Three-level system: theoretical model}
Fig. \ref{HBT}(b) shows a typical $g^{(2)}$ function measured for a single NV at low excitation power. The experimental data are compared with an
equation stemming from a two-level model~\cite{Brouri00,Cuche09}:
\begin{equation}
  g^{(2)}(\tau)=1-e^{-(r+\gamma)\tau}\,,
\label{1}
\end{equation}
where $r$ and $\gamma$ are the excitation and spontaneous emission
rates, respectively. Within this model $g^{(2)}(\tau)<1$, which means that the emitted light is non-classical at any delay $\tau$. However, at higher excitation rate
this simple model (Eq.~\ref{1}) generally fails. This is
particularly true for NV\up{-} centers. In this case, as
shown in Fig. \ref{HBT}(c), the $g^{(2)}$ function
includes a bunching $g^{(2)}(\tau)>1$ feature at finite delays, superimposed to the
antibunching curve. This kind of correlation profile, which
contradicts Eq.~\ref{1}, calls for a third level that traps the electron, preventing
subsequent emission of a photon for a certain
time.~\cite{Doherty1,Doherty2,Maze,Gali,Vincent,Kehayias,Gali09,Felton08}
Therefore, for these delays, the correlation function is higher
than one. Though well known for the
NV\up{-} center, the existence of a shelving level is less
documented for its neutral counterpart but has been invoked
theoretically~\cite{Gali09} as well as in electron-paramagnetic
resonance.~\cite{Felton08} From now on, we will describe phenomenologically the color center dynamics by a three-level model and deduce the intrinsic photo-physical parameters of pure NV\up{-} and NV\up{0} centers (Section IV) and of a photochromic center (section V). These NV centers have been selected from their fluorescence spectra.\\
\subsection{Correlation function of a three-level system}
\begin{figure}[hbtp]
\includegraphics[width=0.7\columnwidth]{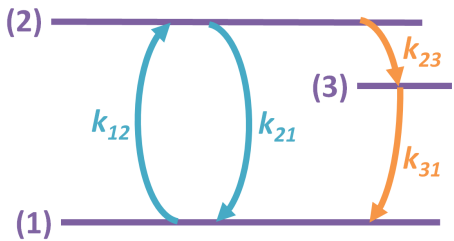}\\
\caption{ (Color online) Jablonski diagram of the three-level system,
including the ground state (1), the excited state (2) and a
metastable state (3). Only the allowed transition channels taken into account in the model are shown.} \label{3niveaux}
\end{figure}

In order to give a quantitative description of the three-level model we
first review briefly the interpretation of the $g^{(2)}$ function
using Einstein's rate equations.  Starting from Glauber quantum
measurement theory \cite{Glauber} we have
\begin{equation}
  g^{(2)}(\tau)=\frac{\langle:I(t+\tau)I(t):\rangle}{(\langle
  I(t)\rangle)^2},\label{3}
\end{equation} where $I(t)$ is the quantum operator equivalent to
the electromagnetic energy flow absorbed by a detector at time $t$
and $::$ represents normal ordering.~\cite{Loudon,Mandel95}
Introducing the creation and annihilation photon operators
$a^{\dagger}(t)$ and $a(t)$ in the Heisenberg representation, we
obtain:
\begin{equation}
  g^{(2)}(\tau)=\dfrac{\langle a^{\dagger}(t)a^{\dagger}(t+\tau)a(t+\tau)a(t)\rangle}{\lvert \langle a^{\dagger}(t)a(t)\rangle \lvert^2 }.
\end{equation}
For a two-level system with ground state $g$ and excited state $e$, the creation operator $a^{\dagger}(t)$ at time
$t$ is to a good approximation proportional to the rising
transition operator $\lvert e,t\rangle \langle g,t\lvert $
\cite{Milonni}, we deduce
\begin{equation}
  g^{(2)}(\tau)=\dfrac{p(e,t+\tau \lvert g,t)}{p(e,t)}=\dfrac{p(e,\tau \lvert
  g,0)}{p(e,0)}
\end{equation} where $p(e,t+\tau \lvert g,t)$ is the conditional
probability for the NV to be in the excited state $e$ at
time $t+\tau$  knowing  that it was in the ground state $g$ at time
$t$. Like in Eq.~\ref{3} this probability is normalized to a single
event probability, i.e., the probability $p(e,t)$ for the NV to
be in the excited state at the previous time $t$.  The
main interest of these equations is to link the probability of
detection, as given by Eqs.~1 and \ref{3}, to the emission
probabilities $p(e,t+\tau
\lvert g,t)$ and $p(e,t)$ defined by the rate equations. Therefore, $g^{(2)}$ can be completely determined if the transient dynamics of the emitter is known.\\
In the context of
the NV center, we must use a two-level system with a third
metastable state to explain the bunching observed in the correlation
measurements.~\cite{footnote2} Since the previous calculations only considered the
excited and ground states involved in the fluorescence process we
will admit (see~\cite{Loudon,Novotny} for a discussion and
justification) that these results still hold with a three-level
system if we replace the excited and ground states by
the levels 1 and 2 in the Jablonski diagram, respectively (see Fig.~3). Here we
neglect the channels 1 to 3 and 3 to 2, because the
system is not supposed to be excited at these transition energies,
in contrast with previous models.~\cite{Kurtsiefer00,Beveratos01b} There, channel 3 to 2 was taken into account and channel 3 to 1 was
neglected, because the quantum yield associated with the NV
relaxation was supposed to be close to unity, while recent studies
show that $Q\simeq 0.6-0.7$.~\cite{Gruber97,Schietinger09,Rittweger,Waldherr,Inam} Therefore, we here take into account four channels. This leads to four unknown parameters: the excitation rate, $ k_{12} $, the
spontaneous emission, $ k_{21} $, and the two parameters $ k_{23} $
and $ k_{31} $ of the additional decay paths involving the shelving level. $ k_{21} $ being the
only radiative channel, we can write the autocorrelation function
as:
\begin{equation}
  g^{(2)}(\tau)=\dfrac{p_{2}(\tau)}{p_{2}(\infty)},\label{gdeux}
\end{equation}
where $ p_2(t)$ is the population of state 2 at time $t$.
$p_{2}(\infty)$ represents the asymptotic limit of $p_2(t)$  when
the transitory dynamics approaches the stationary regime. $ p_2(t)$
can be obtained by solving the system of rate equations defining the
three-level system presented in Fig. \ref{3niveaux}.
With $ p_i $, $i=\{1,2,3\}$, the population of state $ i $ we can write
the following set of equations:
\begin{eqnarray}
\dot{p_1}=-k_{12}p_1+k_{21}p_2+k_{31}p_3,\nonumber\\
\dot{p_2}=k_{12}p_1-(k_{21}+k_{23})p_2,\nonumber\\
\dot{p_3}=k_{23}p_2-k_{31}p_3,\nonumber\\
1=p_1+p_2+p_3,\label{einstein1}
\end{eqnarray}
where $\dot{p_i}$ means time derivative of $p_i(t)$. The use of rate
equations instead of Bloch equations is here fully justified since in ambient conditions, the coherence between levels is decaying very fast.~\cite{Beveratos01b} Eqs.~\ref{einstein1} show that the system is necessarily in one of the states at any time. The steady-state analysis of these equations permits to find the explicit definition
of the fluorescence rate $R$ at which the system emits photons.~\cite{Novotny} This definition allows us to explain the saturation behavior of the NV fluorescence, i.e., $R$ tends towards a finite value for increasing excitation power. However, we are here looking for the time-dependent analysis to find $p_{2}(\tau)$ as a function of the $k_{ij}$ coefficients. If we eliminate $p_3$ from Eq.~\ref{einstein1}, we obtain:
\begin{eqnarray}
\dot{p_1}=-(k_{12}+k_{31})p_1+(k_{21}-k_{31})p_2+k_{31},\nonumber\\
\dot{p_2}=k_{12}p_1-(k_{21}+k_{23})p_2.\label{einstein}
\end{eqnarray}
The resolution of this pair of equations, with the initial conditions
$p_1(0)=1$ and $p_2(0)=0$, leads to $p_{2}(\tau)=p(2,\tau|1,0)$ and
therefore to the expression of $g^{(2)}(\tau)$.\\
Important approximations can be made to obtain a simple
expression. More precisely, in addition to neglecting channels 1 to 3
and 3 to 2, we suppose that :
\begin{equation}
 \lbrace k_{21},k_{12}\rbrace \gg \lbrace k_{23},k_{31}\rbrace.
\end{equation}
This is justified since, even if the third level is considered, the associated rates are supposedly very small compared to the
singlet rates $k_{12}$ and $k_{21}$. We will see latter that this is
not really true for the NV\up{-} center but that the results
obtained are actually robust and keep their meaning even
outside of their \emph{a priori }validity range. Within the above
mentioned approximations the second-order correlation function reads (see Appendix A for mathematical details):
\begin{equation}
  g^{(2)}(\tau)=1-\beta e^{-\gamma_1 \tau}+(\beta-1)e^{-\gamma_2 \tau}\,,
\end{equation}
where the parameters $\gamma_1$, $\gamma_2$ and $\beta$ are defined
through the relations:
\begin{gather}
\gamma_1\simeq k_{12}+k_{21},\\
\gamma_2\simeq k_{31}+\dfrac{k_{12}k_{23}}{k_{12}+k_{21}},\\
\beta\simeq 1+\dfrac{k_{12}k_{23}}{k_{31}(k_{12}+k_{21})}.
\end{gather}
\subsection{Determination of the $k_{ij}$ coefficients}
The aim of this sub-section is to determine the $k_{ij}$ coefficients for the
specific measured NV centers. The fit to the $g^{(2)}(\tau)$ function
allows us to determine $\gamma_1$, $\gamma_2$ and
$\beta$. From Eqs. 11-13 we deduce:
\begin{gather}
k_{21}=\gamma_1-k_{12},\\
k_{31}=\dfrac{\gamma_2}{\beta},\\
k_{23}=\dfrac{\gamma_1 \gamma_2 (\beta-1)}{\beta k_{12}}.
\end{gather}
\begin{center}
\begin{figure*}[hbtp]
\includegraphics[width=2\columnwidth]{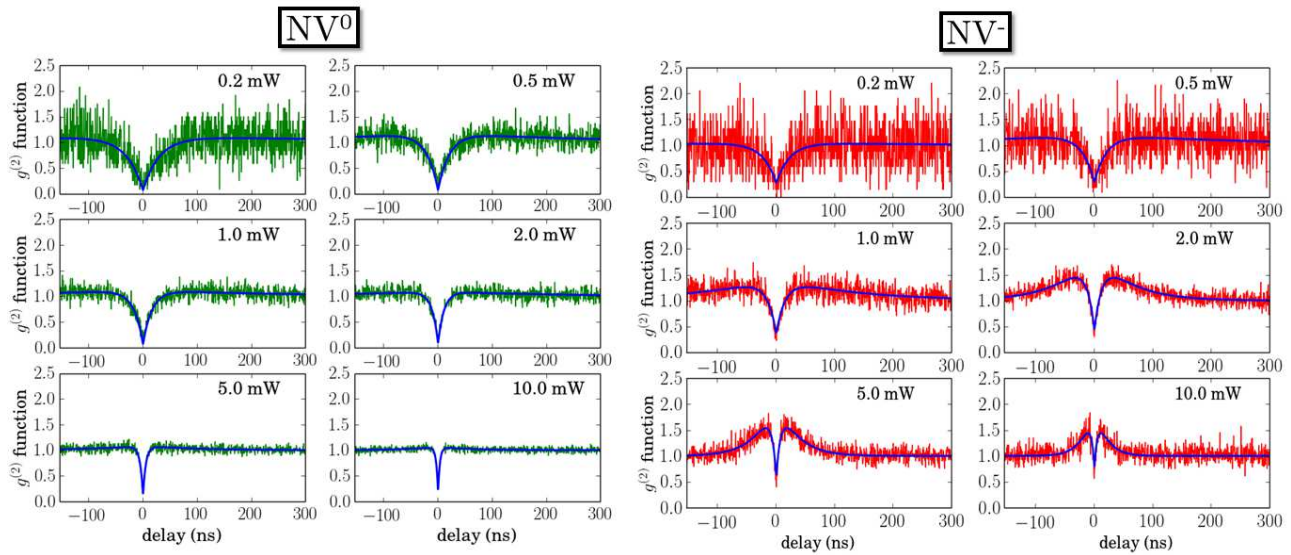}\\
\caption{ (Color online) Time-intensity second-order correlation functions
$g^{(2)}(\tau)$ for a given  NV\up{0} center (green lines) and
NV\up{-} center (red lines). The excitation power is varied from 0.2
mW to 10~mW. The blue lines correspond to a three-level model fit. }
\label{g2}
\end{figure*}
\end{center}
We then have three equations for four unknown variables. A fourth equation is needed to solve the problem entirely. In the experiment, we also have access to the radiation or fluorescence rate $R$, measured in $s^{-1}$. This rate is simply the average number of photons that the APDs collect per second. It represents, up to a multiplicative coefficient associated with the photon propagation in the setup, the probability for the system to be in level 2, multiplied by the transition probability $k_{21}$ to relax (supposedly by optical means) to the ground state. We have:
\begin{equation}
R=\xi k_{21}p_{2}(+\infty),
\label{rate}
\end{equation}
where $\xi$ is the collection efficiency of the system once the NV
center has emitted its fluorescence. Note that the same formula
was used in refs.~\cite{Kurtsiefer00,Beveratos01b} with a different
definition of $p_{2}(+\infty)$ because of a different Jablonski
diagram. However, this has a physical meaning only if
$k_{21}$ is associated with a pure radiative decay. Whereas the model of
ref.~\cite{Kurtsiefer00,Beveratos01b} implied a unity
quantum yield, the quantum yield in our approach is defined by
\begin{equation}
Q=\frac{k_{21}}{k_{21}+k_{23}}.\end{equation} In our
phenomenological approach the third level is thus assumed to absorb
all of the non-radiative transitions letting $k_{21}$ be a pure
radiative decay. Finally we remind that the probability
$p_2(+\infty)$ needed in Eqs.~\ref{rate} and \ref{gdeux} should be
calculated in the asymptotic stationary regime, which can be
obtained by canceling all $\dot{p_i}$ in Eq.\ref{einstein}:
\begin{equation}
p_{2}(+\infty)=\frac{k_{31}}{-k_{21}+k_{31}+(k_{21}+k_{23})(1+k_{31}/k_{12})}.
\end{equation}
Now four equations are at hand so that the system can be inverted to determine all coefficients. This calculation involves the
numerical resolution through Cardano's algorithm~\cite{Cardano} of a third-order
polynomial as given in Appendix B.
\section{Experimental results and extraction of the $k_{ij}$ parameters}

We here consider two representative examples of NDs, one hosting a single pure NV\up{-} center, and the second one hosting a single pure NV\up{0} center. The NV\up{-}-center ND
is about $25~nm$ in diameter, whereas the
one hosting the NV\up{0} center is about $50~nm$ in diameter.~\cite{footnote}  The $g^{(2)}$ function was recorded for both NDs with
different excitation powers $P_{exc}$ in order to study the evolution
of the
$k_{ij}$ coefficients.\\
\begin{figure}[hbtp]
\includegraphics[width=1\columnwidth]{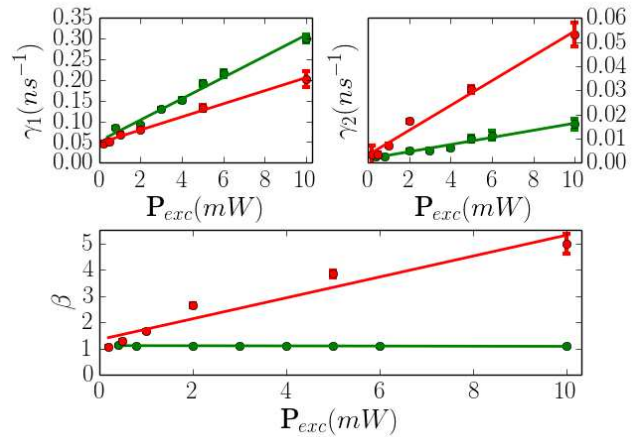}\\
\caption{ (Color online) Evolution of the $\gamma_1$, $\gamma_2$ and $\beta$ parameters as a function of the excitation power. Points are fits to the experimental $g^{(2)}$ functions using Eq. 10 and the lines are linear regressions. The green (red) points and lines correspond to the NV\up{0} (NV\up{-}) center. Errors bars are shown where they exceed the symbol size. They represent the standard deviation of each value  calculated from the covariance matrix as given by the fitting routine.} \label{ParamFit}
\end{figure}
Fig.~\ref{g2} depicts the $g^{(2)}$ function for the two NV centers
and for excitation powers $P_{exc}$ ranging from $200~\mu W$ to
$10~mW$. The experimental curves are fitted with Eq.~10
taking into account the correction for the incoherent background
light collected by the APDs.~\cite{Brouri00,Sonnefraud08,Cuche09} The experimental antibunching
dip does not drop to zero due to this incoherent background, which
modifies Eq.~10 as
$g_{\textrm{exp.}}^{(2)}(\tau)=g^{(2)}(\tau)\rho^2+1-\rho^2$ , where
$\rho=S/(S+B)$ contains the signal $S$ and background $B$
contributions from the NV fluorescence and the spurious
incoherent light, respectively. By recording the average intensity
from the APD directly on the NV and at a location close to it, we
experimentally determine $\rho$ and the fit parameters of
Fig.~\ref{g2} as explained in
refs.~\cite{Brouri00,Sonnefraud08,Cuche09}. It is seen that the
increase of $P_{exc}$ only induces a very small bunching for the
NV\up{0}, contrary to what is observed for the NV\up{-}. However,
the anti-bunching dip narrows with increasing power in both cases.
Our observations are collected in Fig.~\ref{ParamFit}, which shows
the power evolution of the fit
parameters $\gamma_1$, $\gamma_2$ and $\beta$.\\
The general trends seen in Fig.~\ref{g2} are
confirmed in Fig.~\ref{ParamFit} for both NVs since $\gamma_1$, which is associated with the antibunching contribution, is clearly increasing with $P_{exc}$, a fact which is reminiscent from Eq.~2. Furthermore, it is seen that $\gamma_2$ is also increasing significantly for the
NV\up{-} which is clear signature of the third energy level. As far as the $\beta$ parameter is concerned, it
remains at a constant value $\beta\simeq1$ for the NV\up{0}, while it increases up to $\beta\simeq7$ for the NV\up{-}. This behavior agrees with its definition from Eq.~10. Therefore, when $\beta$ is very close to 1, there is no significant bunching, and when $\beta$ increases with power, the bunching turns on.\\
Now that the three parameters have been found, the $k_{ij}$ parameters can be traced back. This will be done in two steps.
\subsection{$k_{ij}$ parameters with constant $k_{21}$}
In Eq.~17 the collection efficiency $\xi$ of the optical setup must be known precisely to extract the various $k_{ij}$. However, since this can only be estimated, we will first calculate the
$k_{ij}$ parameters by assuming that $k_{21}$ does not change with the excitation power. This hypothesis is intuitive because the parameter
$k_{21}$, i.e., the spontaneous emission rate, is supposed to be solely governed by the Fermi's golden rule, which in turn depends on the
electromagnetic environment only. In order to determine the value of
$k_{21}$ we observe that according to Eq.~14 we must have
$\gamma_1=k_{21}$ at zero excitation since in this case $k_{12}=0$. From the linear regression for $\gamma_1$ (Fig.~\ref{ParamFit}) we deduce
$\gamma_1^{(0)}(P_{exc}=0)=k_{21}^{0}=0.052~ns^{-1}$ for the
NV\up{0}, and $\gamma_1^{(-)}(P_{exc}=0)=k_{21}^{-}=0.046~ns^{-1}$
for the NV\up{-} (here the (0) and (-) exponents refer to NV\up{0} and NV\up{-}, respectively). These constants give radiative lifetimes of
$\tau_{21}^{(0)}=19.2~ns$ and $\tau_{21}^{(-)}=21.7~ns$, which are consistent with previous reports (see for example refs.~\cite{Sonnefraud08,Beveratos01b}).\\
\begin{figure}[h]
\includegraphics[width=0.9\columnwidth]{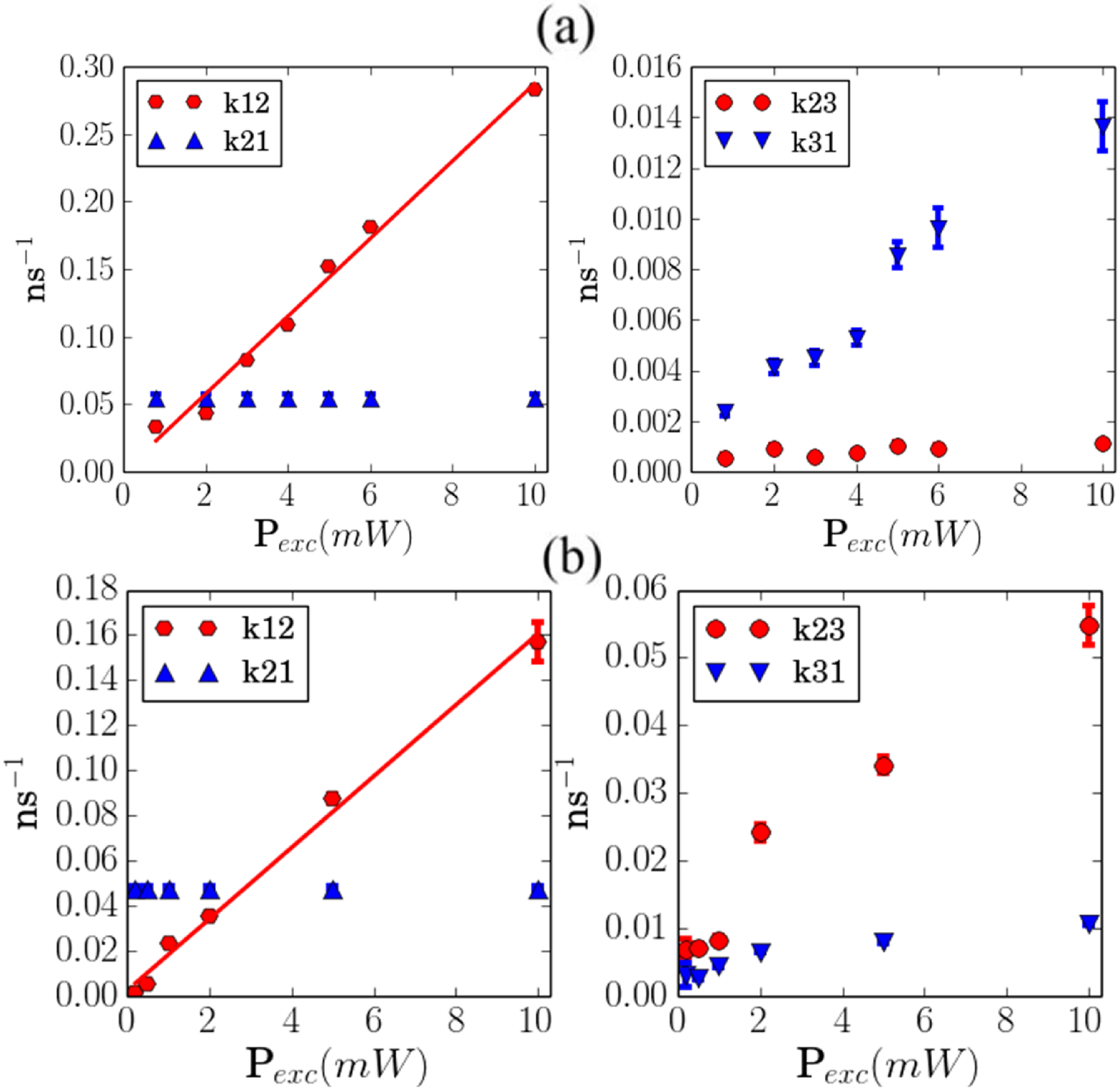}\\
\caption{ (Color online) Evolution of $k_{ij}$ parameters with power excitation
when $k_{21}$ is fixed for (a) the NV\up{0} and (b) for the NV\up{-}. Errors bars are shown where they exceed the symbol size. These error bars are estimated from those of the $\gamma_1$, $\gamma_2$ and $\beta$ parameters, see Eqs. 14 to 16.}
\label{KijBC}
\end{figure}
Thanks to Eqs.~14 to 16, we deduce the three other parameters
as shown in Fig.~\ref{KijBC} for both NVs. The first point to notice is that for both NVs, the $k_{12}$ parameter increases linearly with
the pumping rate from zero to a value exceeding $k_{21}$ for $P_{exc}\simeq
2~mW$. Moreover, for NV\up{0}, we see that $k_{31}$ increases, but keeps very small values compared to the
other parameters, in agreement with the assumptions made in Eq.~9.
However, for NV\up{-}, the same parameters are no longer
negligible compared to the set of $k_{21}$, $k_{12}$ values. They even
overtake $k_{21}$ for $P_{exc}>6~mW$. However, we emphasize that assuming $k_{21}$ constant, if natural, is actually not fully demonstrated. In order to check how robust this hypothesis is, we will now try to approach the values and evolutions of the $k_{ij}$ parameters that were obtained here by modulating the collection efficiency $\xi$ using Eq.~17.
\subsection{Modulation of the collection efficiency $\xi$ }
\begin{figure}[h*]
\includegraphics[width=0.7\columnwidth]{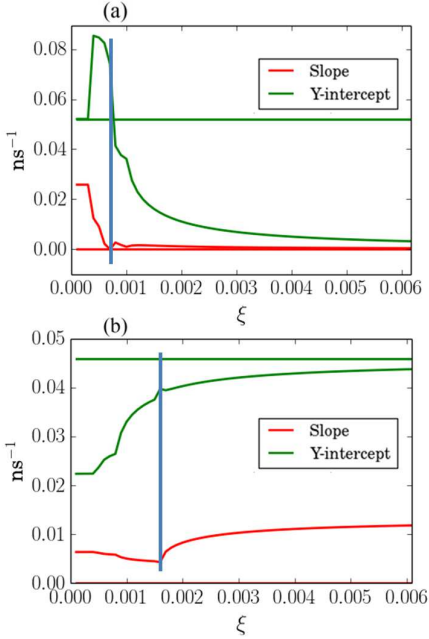}\\
\caption{ (Color online) Evolution of the slope and the Y-intersect of linear fit of
$k_{21}(P_{exc})$ with the collection efficiency $\xi$ for the
NV\up{0} (a) and the NV\up{-} (b). Straight lines are the values
obtained previously with $k_{21}$ fixed.} \label{EffBC}
\end{figure}

Now we use the fourth equation Eq.~17 to calculate the $k_{ij}$ parameters. As already stated, we
do not know exactly the $\xi$ parameter but, as it turns out, slight variations  in $\xi$ can produce significant changes in $k_{ij}$. To find the correct value of $\xi$, we adopt the following procedure. We let its value vary continuously and calculate the evolutions of the slope and Y-intercept of the linear regressions made with the obtained $k_{21}(P_{exc})$ traces. The slope should vanish because $k_{21}$ is assumed to be constant, whereas the Y-intercept should reach the value obtained previously, i.e., the value of $\gamma_1$ at zero excitation. Therefore, we calculate the evolution of the
$k_{ij}$ parameters as a function of the collection efficiency
$\xi$. The results are shown in Fig.~\ref{EffBC}, where the constant horizontal curves depict the values to be reached by the slope and Y-intercept. For the NV\up{0} (Fig.~\ref{EffBC} (a)), there is indeed a value of $\xi$ where the two parameters reach the assumed values (blue vertical line). This gives  $\xi^{(0)}=0.77\times 10^{-3}$, in agreement with a rough estimate of our setup collection efficiency taking into account the various optical components.
However, for the NV\up{-} (Fig.~\ref{EffBC} (b)), it is seen that
the parameters do not reach exactly the previous values. Yet, there
is an optimum $\xi$ for which the $k_{ij}$ approach the previous
values (blue vertical line). This corresponds to
$\xi^{(-)}=1.6\times 10^{-3}$, which differs only by a factor 2 from
the NV\up{0} case. This appears reasonable since the measurements
were not carried out the same day (optical alignments might be slightly different) and the
collection efficiency $\xi$ also depends on the
unknown transition-dipole orientation in both NVs.
\subsection{Comparaison of the photophysics of NV centers in both charge states}
\begin{figure}[h*]
\includegraphics[width=0.9\columnwidth]{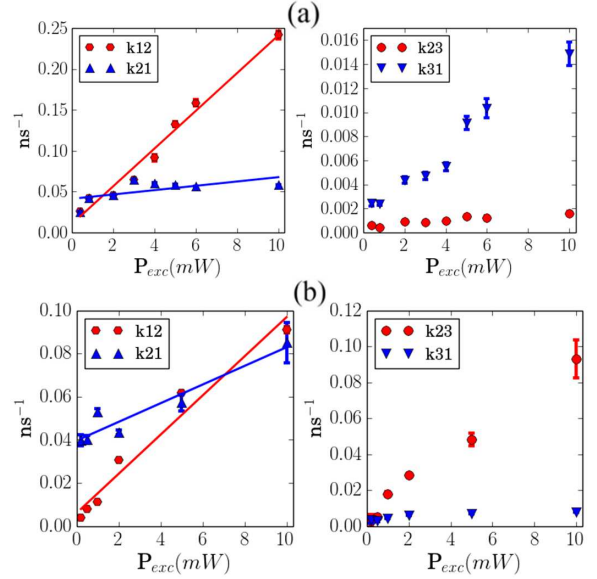}\\
\caption{ (Color online) Evolution of the $k_{ij}$ parameters  with  the
excitation power taking into account the collection efficiency for the
NV\up{0} (a) and the  NV\up{-} (b) centers. Errors bars estimated as in Fig.~\ref{KijBC} are shown where they exceed the symbol size.} \label{KijBCm}
\end{figure}

Fig.~\ref{KijBCm} depicts the $k_{ij}(P_{exc})$ curves deduced
from the previous optimization. It is found that the
order of magnitude of the coefficients is the same as
for imposed $k_{21}$. In particular, the relaxation rate
$k_{31}$ is unchanged because it only depends on
the $g^{(2)}$ fit parameters, Eq.~15. For the two NV centers, the
excitation rate vanishes in the absence of any excitation power and
linearly increases with $P_{exc}$ as it should. The main difference between  both centers comes from the evolution of $k_{23}$, $k_{31}$ and $k_{21}$. Indeed,
for NV\up{0}, $k_{23}$ and $k_{31}$  remain very small compared
to $k_{21}$, which is almost constant ($1/k_{31}\approx 500~ns$ and
$1/k_{23}\approx 1000~ns$ if $P_{exc}$ tends to zero). Therefore, the third level plays little role in the photodynamics of the NV\up{0} center. However, it is worth stressing that although very small, $k_{23}$ is not zero for NV\up{0} (errors bars are within the symbol size in Fig.~\ref{KijBCm} (a), right panel), which confirms the very existence of the third metastable level for this charge state. Regarding the $g^{(2)}$ curves, it
implies that the optical channel 2 to 1 is favored, which prevents
any significant bunching. Furthermore, for the NV\up{0} the
narrowing of the antibunching dip is due to
the increasing excitation rate as for a two-level system, i.e., Eq.~2 (see ref.~\cite{Cuche09}).\\
The analysis of the $k_{ij}$ parameters
for the NV\up{-} is more involved. Indeed, $k_{23}$ increases
very quickly to reach the order of magnitude of $k_{21}$, while the
latter is increasing as well (at zero excitation power we have
$1/k_{21}\simeq 24$ ns and $1/k_{23}\simeq 500$ ns). The increase of
$k_{23}$, associated with non-radiative transitions, actually
explains the growth of the bunching feature on the $g^{(2)}$ curves. Although the physical justification of this finding is beyond our
phenomenological treatment, it is likely that the variation of $k_{21}$ and $k_{23}$
with excitation power is due to a change in the local
energy environment of the NV center at high power, in particular
because the efficient coupling with the phonon bath in the diamond matrix is
expected to be temperature sensitive.
\begin{figure}[h*]
\includegraphics[width=0.9\columnwidth]{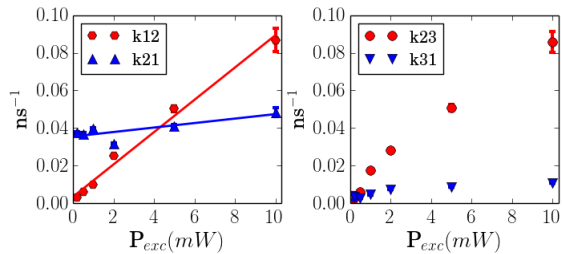}\\
\caption{ (Color online) Evolution of $k_{ij}$ parameters with power excitation
taking into account the collection efficiency for the  NV\up{-} in
the approximation-free rate-equation model. Errors bars estimated as in Fig.~\ref{KijBC} are shown where they exceed the symbol size.} \label{trucmuch}
\end{figure}
Increasing the excitation power could thus correspond to an increasing effective temperature, subsequently affecting the relaxation dynamics.\\
It is worth pointing out
that there is a limitation in the
analysis done for the NV\up{-} center. Indeed, the very fact
that $k_{23}$ and $\lbrace k_{21},k_{12}\rbrace$ reach the same order
of magnitude contradicts the hypothesis made in Eqs.~9-13.
Actually, as already mentioned, the results obtained are much more
robust that could be anticipated at first sight. This can be figured out by relaxing the constraint of
Eq.~9 as done in the detailed calculation presented in Appendix C. The results obtained with the new rate-equation model are shown in Fig.~\ref{trucmuch} for
\begin{figure}[h*]
\includegraphics[width=5 cm]{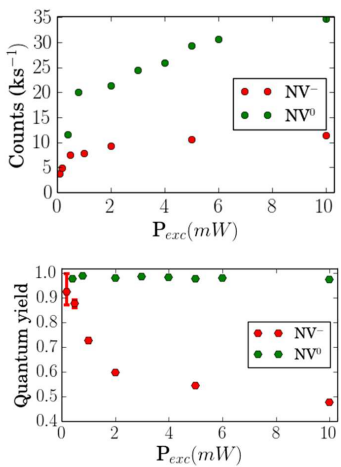}\\
\caption{ (Color online) Evolution of the fluorescence rate (upper panel) and the quantum yield $Q$ (lower panel) with the excitation power for the NV\up{-} and NV\up{0} center in the rate equation
model free of any approximation. Errors bars for $Q$ are shown where they exceed the symbol size. They are estimated from those of the $k_{ij}$ parameters, see Eq. 18.} \label{retrucmuch}
\end{figure}
the evolution of the $k_{ij}$ coefficients of the NV\up{-} center. The obtained values are very
similar to those of the approximate treatment, thereby
justifying the previous results. For the NV\up{0} the
coupling to the third level is very weak and the modifications (not
shown) are even smaller.\\
To complete the analysis we also computed the
quantum yield evolution as given by Eq.~18 and compared with the
evolution of the fluorescence rate. Fig.~\ref{retrucmuch} confirms
that within the Jablonski model sketched in Fig.~3 and in the considered excitation
regime, the quantum yield of the NV\up{0} center is approximately constant $Q\simeq1$, in agreement with the intuitive fact that the third level does not play a significant role
in the dynamics. In contrast, the NV\up{-} quantum yield decreases
dramatically with increasing excitation power from a starting $Q\simeq 1$
to $Q\simeq 0.5$ at high power. This entails the fact that the
NV\up{-} dynamics is strongly dependent on the excitation power as
discussed before.

\section{Photochromism}

\begin{figure}[h*]
\includegraphics[width=0.9\columnwidth]{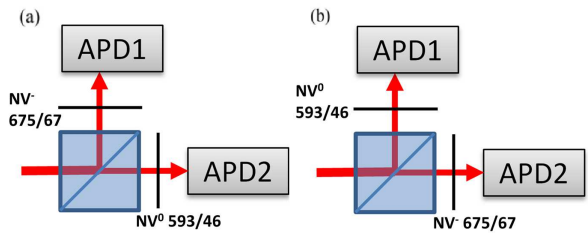}\\
\caption{ (Color online) HBT configurations used to study photochromism. (a) is the NV\up{-/0} configuration, where a bandpass
filter (central wavelength 675~nm, bandwidth 67~nm) adapted to the negative NV\up{-} center is inserted in the
upper start branch, whereas a bandpass filter adapted to the neutral
NV\up{0} (central wavelength 593~nm, bandwidth 46~nm) is inserted in the lower stop branch. (b) is the NV\up{0/-}
configuration where both bandpass filters are interchanged compared
to (a).} \label{HBTNV}
\end{figure}

Photochromism of NV centers has been
reported in ensembles of NV centers in CVD diamond films under
additional selective illumination~\cite{Iakoubovskii00}, with
single NV centers in 90~nm NDs under femtosecond illumination,
which results in the photo-ionization of the negative center to its
neutral counterpart~\cite{Dumeige04}, with ensembles of NV centers
in type Ib bulk diamond at cryogenic temperatures under intense CW
excitation~\cite{Manson05}, and with a single center in natural
type-IIa bulk diamond under CW illumination.~\cite{Gaebel06} In this
last report, a special scheme of cross-correlation photon
measurements was applied in the emission band of both charge states
to show that the collected fluorescence in the NV\up{0} and NV\up{-}
states were correlated and originated form a single NV defect. Several studies have reported that charge conversion within the NVs critically depends on the illumination conditions.~\cite{Waldherr,beha,aslam} \\
A complete understanding of NV photochromism is lacking but a
widespread view is that the optical excitation, either CW or
transient, tunes the quasi Fermi level around a NV charge transition
level, thereby inducing charge conversion.~\cite{Iakoubovskii00,Grotz12} This scenario has been
reinforced recently by electrical manipulation of the charge state
of NV ensembles and of single NVs by an electrolytic gate electrode
used to tune the Fermi energy.~\cite{Grotz12} Depending on the sort
of diamond studied, photochromism is thought to be favored by the
presence of electron donor or acceptor defects, such as nitrogen, in
the neighborhood of the NV center.~\cite{Collins02} Recently, it was
also found that resonant excitation of the NV\up{0}  and NV\up{-}
states in ultrapure synthetic IIa bulk diamond can induce reversible
charge conversion in cryogenic conditions even at low
power.~\cite{Siyushev13} This was taken as evidence that the charge
conversion process is intrinsic in this sort of diamond, not
assisted by an electron donor or acceptor state. The goal of this
section is to give additional information on NV photochromism
detected in surface-purified NDs, 25~nm in size, subjected to a CW
non-resonant excitation of increasing power. By comparing
the behavior of a single photochromic center to that
of non-photochromic centers in the same illumination conditions as described above, we gain
valuable information on the rich photophysics of photochromism.\\
For the
purpose of studying photochromism, we use two additional
configurations of the HBT correlator that differ only by the set of
bandpass filters added in the interferometer branches. These
configurations, called NV\up{-/0}, and NV\up{0/-} respectively, are
shown schematically in Fig.~\ref{HBTNV}.
In contrast to Fig.~\ref{HBT}(a), these two configurations add
selective bandpass filters in the interferometer branches. The
NV\up{-/0} configuration in Fig.~\ref{HBTNV} (a) (respectively
NV\up{0/-} configuration in Fig.~\ref{HBTNV} (b)) uses a filter
selective to the NV\up{-} (respectively NV\up{0}) fluorescence in
the start (respectively stop) branch. Therefore, in the NV\up{-/0}
configuration, a single photon emitted by a NV\up{-} center and detected
in APD1 gives the \og Start \fg{} signal to the counting module,
whereas a single NV\up{0} photon subsequently detected in APD2
produces the \og Stop \fg{} signal. The NV\up{0/-} configuration works in
just the complementary way. Note that our NV\up{-/0} configuration is
similar to the cross-correlation technique used in ref.
~\cite{Gaebel06}. These schemes turn out to be very powerful to
study photochromism since cross correlation can be expected only if
NV\up{-} and NV\up{0} photons originate from the same defect center.
Note that related techniques have also successfully been
applied to identify various excitonic species emitted by single
semiconductor quantum
dots, see for instance.~\cite{Moreau01,Regelman01,Kiraz02,Couteau04,Sallen}\\

\begin{figure}[h*]
\includegraphics[width=1\columnwidth]{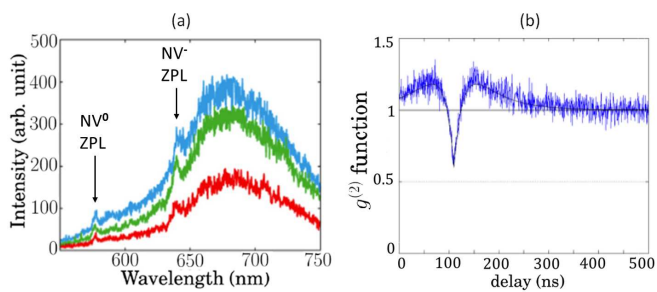}\\
\caption{ (Color online) (a) Fluorescence spectra of the photochromic ND nanoparticle taken at 3
different excitation powers of nominally 0.5~mW, 1~mW, 2~mW from bottom to top, respectively. (b) Antibunching curve at short time delays and an excitation power of 5~mW. }
\label{Spectres}
\end{figure}

With the above setup, we have located a few rare NDs
hosting a single NV center that showed charge
conversion at low excitation power already. In the following, we
consider such a ND hosting a single photochromic NV center.

The relevant spectra are shown in Fig.~\ref{Spectres}(a). It is
found that both the NV\up{0} and NV\up{-} ZPLs are seen at any
excitation power. The corresponding $g^{(2)}$ function measured in
the standard configuration of the HBT (no bandpass filter) is shown
in Fig.~\ref{Spectres}(b). It reveals a clear antibunching dip at
zero delay. A precise analysis taking into account the background light (see Table 1)
confirms this finding since at high excitation power we observe that
the background $B$ increases significantly with respect with the NV
fluorescence signal:  $\rho=S/(S+B)\simeq 0.6$ while at low power
$\rho\simeq 0.9-1$. With this value and using the formula
$g_{\textrm{exp.}}^{(2)}(\tau)/\rho^2+1-1/\rho^2=g^{(2)}(\tau)$ we
deduce the actual value of $g^{(2)}(0)\simeq 0$.~\cite{Beveratos01,Sonnefraud08,Beveratos01b,Hui09} From these results, a natural interpretation for the observation of the NV\up{0} and
NV\up{-} ZPLs together with NV uniqueness is that this particular ND
is subjected to photochromism. In addition to the antibunching dip
at zero delay, it is seen in  Fig.~\ref{Spectres} that $g^{(2)}$
exceeds 1 at longer delays.~\cite{Beveratos01b} In agreement with the
previous sections, this is evidence for the presence of a trapping
level and calls for a three-level description of the photochromic
NV. The values of $\gamma_1$, $\gamma_2$ and $\beta$ parameters used for
fitting the $g_{\textrm{exp.}}^{(2)}$ function agree qualitatively
well with those obtained for the NV\up{-} considered previously in
agreement with the fact that the system is acting more
like a three-level system.\\
Although our study of NV photochromism is limited to a particular example we suggest that the dynamics of the system involves probably all energy levels of the NV\up{-} and NV\up{0} with some possible hybridization. It could be interesting to know, whether or not, the third level involved in the photochromic case is identical in nature to the third level of the NV\up{-}. The role of the environment or of the radiation power~\cite{TreussartPhysicaB2006} on the dynamics could be investigated in the future to clarify this point.
\\
\begin{table}
\begin{tabular}{c|c|c|c|c|c}
 $P_{exc.}$(mW) &0.5 &1&2&3&5\\ \hline\hline
 $R$(kHz)& 11.8&17.1&23.0&27.7&32.0\\ \hline
 $g_{\textrm{exp.}}^{(2)}(0)$&0.25&0.32&0.48&0.45&0.6\\ \hline
 $S$(kHz)&10.2&14.1&16.6&20.5&20.2\\ \hline
 $B$(kHz)&1.6&3.0&6.4&7.2&11.8\\ \hline
 $\beta$&1.36&1.45&1.75&1.6&2.5\\ \hline
 $\gamma_1$(ns$^{-1}$)&0.03&0.036&0.044&0.052&0.054\\ \hline
 $\gamma_2$(ns$^{-1}$)&0.005&0.007&0.009&0.012&0.018 \\
\hline\hline
\end{tabular}
\caption{Table summarizing the experimental parameters $P_{exc.}$,
$R$, $g_{\textrm{exp.}}^{(2)}(0)$, $S$, $B$ measured on the
photochromic NV for the observation of the $g_{\textrm{exp.}}^{(2)}$
function. The fit parameters $\gamma_1$,$\gamma_2$ and
$\beta$ using a 3 energy-level model are also given. }
\end{table}
\begin{figure}[h*]
\includegraphics[width=1\columnwidth]{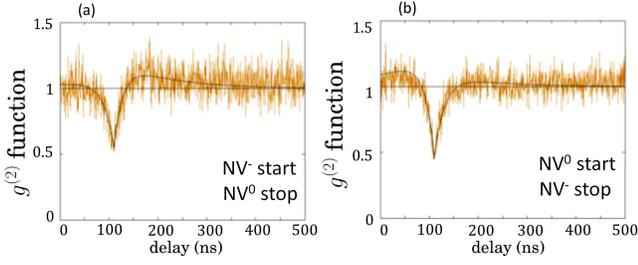}\\
\caption{ (Color online) Time-intensity second-order correlation functions measured
for the photochromic NV center in two different HBT
configurations. (a) is the cross correlation NV\up{-/0}
configuration; (b) is the cross correlation NV\up{0/-}
configuration. The temporal window is [-100~ns;
500~ns] and the excitation power is 3~mW.} \label{Fig7}
\end{figure}
Switching now to the cross-correlation regime we show in Fig.~\ref{Fig7} the result obtained  for
$P_{exc.}=3~mW$ (similar features, not shown here, were observed at
other excitation powers). It is worth emphasizing that cross correlations, as
described in e.g. refs.~\cite{Aspect,Regelman01,Kiraz02}, allow us
to characterize the transitory dynamics between the two NV
configurations. Using a formalism equivalent to the one leading to
Eqs.~3-5 we indeed obtain in the NV\up{-/0} configuration sketched
in  Fig.~\ref{HBTNV} (a)
\begin{equation}
  g^{(2)}_{-/0}(\tau)=\dfrac{p(e, \textrm{NV\up0},t+\tau \lvert
  g,\textrm{NV\up-},t)}{p(e,\textrm{NV\up-},t)}\label{bbb}
\end{equation} where $p(e, \textrm{NV\up0},t+\tau \lvert g,\textrm{NV\up-},t)$ is the conditional
probability for the NV to be in the excited level $e$ of
the NV\up0 state at time $t+\tau$  knowing  that it was in the
ground energy level $g$ of the  NV\up- charged state at the previous
time $t$. This also means that a first photon emitted by the
NV\up0 was detected at time $t$ while a second photon emitted by the NV\up- is detected at time $t+\tau$ . In a symmetrical way using the
NV\up{0/-} configuration sketched in  Fig.~\ref{HBTNV} (b) we get
\begin{equation}
  g^{(2)}_{0/-}(\tau)=\dfrac{p(e, \textrm{NV\up-},t+\tau \lvert g,\textrm{NV\up0},t)}{p(e,\textrm{NV\up0},t)}
\label{bbbb}
\end{equation} with similar definitions as previously but with the
role of NV\up0 and NV\up- inverted.\\
The experimental results corresponding to these two configurations
are shown in Figs.~\ref{Fig7}(a) and (b), respectively. Here, it is
also important to have $\tau\geq0$ in the calculation leading to
Eqs.~\ref{bbb} and \ref{bbbb} in order to have a clear physical
understanding. However, the electronic delay $\Delta=100$ ns
included in the HBT correlator setup implies that sometimes even a
photon emitted, say at time $t_0$, by a NV\up- is recorded after a
second photon emitted later by the NV\up0 state, i.e.,  at
$t_0+\tau$. This corresponds to the 'negative' delay part of the
graph, i.e., Fig.~\ref{Fig7}(a), which is actually associated with
the inverse dynamics NV\up{0/-}, i.e., Eq.~\ref{bbbb}. Here for
clarity we did not subtract the delay from the abscises in
Fig.~\ref{Fig7}. Therefore for $\tau>\Delta$ in Fig.~\ref{Fig7}(a)
we have
$g^{(2)}_{exp.}(\tau)=g^{(2)}_{-/0}(\tau-\Delta)\rho^2+1-\rho^2$
while we have
$g^{(2)}_{exp.}(\tau)=g^{(2)}_{0/-}(\Delta-\tau)\rho^2+1-\rho^2$ for
$0<\tau<\Delta$ with $\rho=S/(S+B)$ as previously. In the same way
for Fig.~\ref{Fig7}(b) we have
$g^{(2)}_{exp.}(\tau)=g^{(2)}_{0/-}(\tau-\Delta)\rho^2+1-\rho^2$ for
$\tau>\Delta$ and
$g^{(2)}_{exp.}(\tau)=g^{(2)}_{-/0}(\Delta-\tau)\rho^2+1-\rho^2$ for
$0<\tau<\Delta$.\\
As it is clear from the definitions these
cross-correlations should present some symmetries. In the present
case we used a three energy-level fit, i.e. Eq.~10, for the
theoretical functions, Eqs.~\ref{bbb} and \ref{bbbb}. The parameters
obtained to reproduce the data  are the same for
Fig.~\ref{Fig7} (a) and (b) up to an inversion between the
`positive' and `negative' delay for each graph. Therefore, we get
for Fig.~\ref{Fig7}(a) $\gamma_1=0.046~ns^{-1} $, $\gamma_2=0.01~ns^{-1}$ and $\beta=
1.2$ for $\tau<\Delta$, while we have $\gamma_1=0.052~ns^{-1}$,
$\gamma_2=0.01~ns^{-1}$ and $\beta= 1.5$ for $\tau>\Delta$. For the second
cross-correlation curve the parameters are identical but the roles of
$\tau>\Delta$ and $\tau<\Delta$ are inverted as it should be. We
observe that these values are very close to each other and also
from the one obtained in Table 1 at the same excitation power
$P_{exc.}=3~mW$. This confirms that the system acts here mainly as a
NV\up- center.\\
Interestingly, the last finding implies that after the
emission of a photon in the spectral fluorescence band  of the
NV\up- the delayed emission of a second photon in the spectral band
of the NV\up0, i.e., the conditional probability given by
Eq.~\ref{bbb}, is also characterized by the dynamics of the NV\up-
contrary to the intuition. Such behavior was reported in
ref.~\cite{Gaebel06} for NV centers in bulk. In particular, in this
paper the time dependence of the conversion NV\up0 to NV\up- process
(and its inverse) was studied using pulse sequences. It was found
that the relaxation from NV\up- to NV\up0 is a very slow process
occurring with a decay time $\simeq 1~ \mu s$. This agrees with our
finding in Fig.~\ref{Fig7} since, even if the full dynamics of the
photochromic NV center is expected to depend on the energy levels of
both the NV\up- and NV\up0 centers, it is clearly the NV\up-
characteristics which dominate during the transition associated with
Eq.~\ref{bbb}. A similar qualitative analysis can be done in the
NV\up0 to NV\up- center conversion. The small dissymmetry between
the two parts of the curves for positive and negative delays results
from the presence of a small NV\up0 contribution to the dynamics
during the NV\up0 to NV\up- transition which is absent in the NV\up-
to NV\up0 conversion. Clearly, this complex charged/uncharged
transition dynamics would deserve systematic studies in the future.
\section{Summary}
To summarize, we have experimentally studied the
fluorescence photodynamics of NV\up- and NV\up0 centers in diamond nanocrystals of 50~nm size or below using HBT
photon-correlation measurements as a function of the excitation power. The dynamics was theoretically modeled using Einstein's
rate equations and the transition probability rates $k_{ij}$
entering the three-level model developed to analyze the data were deduced and used to infer a quantum efficiency to both charge states of the NV. It has been found that the shelving state, though present, plays a very small role on the neutral center in those small diamond crystals. The narrowing of the antibunching dip observed with increasing power for this center is a simple power effect that does not affect the near unity quantum efficiency. In contrast, the negative center experiences a distinctive photon bunching behavior at finite delay that increases with increasing power. This reflects the increasing role of the shelving state for this center, which in turn diminishes the quantum efficiency from near unity at low power to approximately 0.3 at high power. We have also studied the dynamics of a photochromic center. It reveals a rich phenomenology that is essentially dominated by the negative face of the center. In the future, it would be interesting to determine how the presence of, e.g., a plasmonic structure affects the dynamics of NV centers~\cite{Schietinger09} in any charge state, in particular if the role of the shelving state in the NV\up0 center as well as the neutral face of the photochromic center can be modified.\\
\section{Acknowledgments}
This work was supported by Agence Nationale de la Recherche, France, through the NAPHO (grant ANR-08-NANO-054-01), PLACORE (grant ANR-13-BS10-0007) and SINPHONIE (grant ANR-12-NANO-0019) projects.

\appendix
\section{}

Here we briefly summarize the derivation of the main equations used
in sections III and IV.  We start from Eq.~8 written in a matrix form
as
\begin{eqnarray} \dot{\mathbf{P}}:=\left(\begin{array}{c}\dot{p_1}\\ \dot{p_2}
\end{array}\right)=\left(\begin{array}{cc} a& b
\\ c& d
\end{array}\right)\cdot \mathbf{P} +\mathbf{J}\nonumber\\=\mathbf{M} \cdot \mathbf{P} +\mathbf{J},\label{diff}
\end{eqnarray}
where
\begin{eqnarray}a=
-(k_{12}+k_{13}+k_{31}),\nonumber\\b=k_{21}-k_{31},\nonumber\\c=k_{12}-k_{32},\nonumber\\d=-(k_{21}+k_{23}+k_{32}),\nonumber
\end{eqnarray}and
\begin{eqnarray} \mathbf{J}=\left(\begin{array}{c}k_{31}\\ k_{32}
\end{array}\right).
\end{eqnarray}
For the sake of generality we here keep all transition coefficients
$k_{ij}$ allowed by the three-level model. Such an Eq.~\ref{diff}
can be formally solved by defining a new vector $\mathbf{q}$ related
to $\mathbf{P}$ by the equation
$\mathbf{P}=e^{\mathbf{M}t}\cdot\mathbf{q}$.  This leads to the new
equation $\dot{\mathbf{q}}=e^{-\mathbf{M}t}\mathbf{J}$ which has the
general solution:
\begin{eqnarray}
\mathbf{q}(t)=\mathbf{q}(0)+\int_0^t d\tau
e^{-\mathbf{M}\tau}\cdot\mathbf{J}\nonumber\\
=\mathbf{q}(0)+\mathbf{M}^{-1}\cdot[1-e^{-\mathbf{M}t}]\cdot\mathbf{J}\label{solution}
\end{eqnarray} where $\mathbf{M}^{-1}$ is the inverse matrix of $\mathbf{M}$ and where the initial condition
corresponding to a system in the ground state at time $t=0$ is  $\mathbf{q}(0)=\mathbf{P}(0)=\left(\begin{array}{c}1\\
0
\end{array}\right)$.  Therefore, we deduce:
\begin{eqnarray}
\mathbf{P}(t)
=e^{\mathbf{M}t}\cdot(\mathbf{q}(0)+\mathbf{M}^{-1}\cdot[1-e^{-\mathbf{M}t}]\cdot\mathbf{J})\label{solution}
\end{eqnarray}
In order to give an explicit form to the solution we need to diagonalize the $\mathbf{M}$ matrix. Therefore, we write
\begin{eqnarray}
\mathbf{\Pi}^{-1}\cdot\mathbf{M}\cdot\mathbf{\Pi}=
\left(\begin{array}{cc} \lambda_1& 0
\\ 0& \lambda_2
\end{array}\right).
\end{eqnarray}
The eigenvalues $\lambda_1$ and $\lambda_2$  of $\mathbf{M}$ are
easily obtained from  the secular equation $det[\mathbf{M-\lambda
\mathbf{I}}]=0$. They reads:
\begin{eqnarray}
\lambda_1=\frac{a+d}{2}-\frac{1}{2}\sqrt{[(a+d)^2-4det[\mathbf{M}]]}\nonumber\\
\lambda_2=\frac{a+d}{2}+\frac{1}{2}\sqrt{[(a+d)^2-4det[\mathbf{M}]]}.\label{value}
\end{eqnarray}
From Eq.~\ref{solution} we easily obtain
\begin{eqnarray}
\mathbf{\Pi}^{-1}\cdot\mathbf{P}= \left(\begin{array}{cc}
e^{\lambda_1 t}& 0
\\ 0& e^{\lambda_2 t}
\end{array}\right)\left(\begin{array}{c}\alpha\\ \varepsilon
\end{array}\right)+\left(\begin{array}{c}\gamma\\ \delta
\end{array}\right),\label{ouais}
\end{eqnarray} with
\begin{eqnarray}
\left(\begin{array}{c}\gamma\\ \delta
\end{array}\right)=-\left(\begin{array}{cc} 1/\lambda_1& 0
\\ 0& 1/\lambda_2
\end{array}\right)\cdot\mathbf{\Pi}^{-1}\cdot\mathbf{J}
\end{eqnarray}
and
\begin{eqnarray}
\left(\begin{array}{c}\alpha\\ \varepsilon
\end{array}\right)=\mathbf{\Pi}^{-1}\cdot\left(\begin{array}{c}1\\ 0
\end{array}\right)-\left(\begin{array}{c}\gamma\\ \delta
\end{array}\right).\label{const}
\end{eqnarray}
In order to determine completely the solution we need to specify the
transformation matrix $\mathbf{\Pi}=\left(\begin{array}{cc} u_1& u_2
\\v_1& v_2
\end{array}\right)$ whose column vectors $\mathbf{X}_i=\left(\begin{array}{c}u_i\\
v_i
\end{array}\right)$ with $i=1,2$ are solutions of the eigenvalue
problem $\mathbf{M}\cdot\mathbf{X}_i=\lambda_i \mathbf{X}_i$. These
eingenvectors are determined up to an arbitrary normalization and
here we choose
\begin{eqnarray} \mathbf{X}_i=\left(\begin{array}{c}b\\ \lambda_i-a
\end{array}\right).\end{eqnarray}
Using  Eq.~\ref{ouais} we therefore get:
\begin{eqnarray}
p_2(t)=v_1\alpha e^{\lambda_1 t}+v_2\varepsilon e^{\lambda_2
t}+v_1\gamma+v_2\delta.
\end{eqnarray}
We can further simplify the solution since from Eq.~\ref{const} and
the definition of $M^{-1}$ we easily deduce $(v_1 v_2)\left(\begin{array}{c}\alpha\\
\varepsilon
\end{array}\right)=-(v_1 v_2)\left(\begin{array}{c}\gamma\\ \delta
\end{array}\right)$ , i.e., $v_1\alpha
+v_2\varepsilon=-(v_1\gamma+v_2\delta)$. After regrouping all terms we obtain
\begin{eqnarray}
\frac{p_2(t)}{p_2(\infty)}=-\frac{v_1\alpha}{v_1\alpha
+v_2\varepsilon} e^{\lambda_1 t}+(\frac{v_1 \alpha}{v_1\alpha
+v_2\varepsilon}-1)
e^{\lambda_2 t}+1\nonumber\\
\end{eqnarray} which up to notations is equivalent to Eq.~10 of Section III if we write $\lambda_i=-\gamma_i$ and $\beta=\frac{v_1\alpha}{v_1\alpha
+v_2\varepsilon}$.
\section{}
The results obtained in Appendix A are exact and no approximation were
made for calculating the coefficients $\gamma_i$ and $\beta$. Now we
will use the fact that $\lbrace k_{21},k_{12}\rbrace \gg \lbrace
k_{23},k_{31}\rbrace$ to simplify and explicit these
coefficients.\\
First,  we point out that we have
\begin{eqnarray}
\beta=\frac{(1-J_1/\gamma_1)}{(\gamma_1-\gamma_2)}\cdot\frac{\gamma_1\gamma_2}{J_1}\nonumber\\=\frac{(1-k_{31}/\gamma_1)}{\sqrt{((a+d)^2-4det[\mathbf{M}])}}\frac{det[\mathbf{M}]}{k_{31}}.
\label{co}
\end{eqnarray}
Therefore from Eqs.~\ref{value} and \ref{co} we see that all
coefficients $\gamma_i$ and $\beta$ can be expressed as functions of
$a+d$ and $det[\mathbf{M}]$. Up to the first order we have
\begin{eqnarray}
a+d\simeq-(k_{12}+k_{21}),\nonumber\\
det[\mathbf{M}]\simeq k_{31}(k_{12}+k_{21})+ k_{23}k_{12},
\end{eqnarray}
Therefore, up to the same order, we have
$\sqrt{((a+d)^2-4det[\mathbf{M}])}\simeq
-(a+d)[1-4det[\mathbf{M}]/(a+d)^2]$.  These lead to
\begin{eqnarray}
\gamma_1\simeq -(a+d)=k_{12}+k_{21},\nonumber\\
\gamma_2\simeq-\frac{det[\mathbf{M}]}{a+d}\simeq
k_{31}+\frac{k_{23}k_{12}}{k_{12}+k_{21}},
\end{eqnarray}  which are  Eqs.~11 and 12, respectively. Finally we
have
\begin{eqnarray}
\beta\simeq \frac{-1}{(a+d)}\frac{det[\mathbf{M}]}{k_{31}}\simeq
1+\frac{k_{23}k_{12}}{k_{31}(k_{12}+k_{21})}
\end{eqnarray}  which is Eq.~13.\\
In order to solve the system of equations 11-13 we first eliminate
$k_{23}$ from Eq.~12 using Eq.~13, i.e.
\begin{eqnarray}
k_{23}=\dfrac{\gamma_1 k_{31}(\beta-1)}{k_{12}}.\label{truc}
\end{eqnarray}
Inserting this result into Eq.~11 leads to
\begin{eqnarray}
k_{31}=\dfrac{\gamma_2}{\beta}
\end{eqnarray} which constitutes our first parameter solution. In
order to determine the other parameters we insert the value obtained for
$k_{31}$ into Eq.~\ref{truc} and Eq.~11 to obtain
\begin{eqnarray}k_{23}=\dfrac{\gamma_1
\gamma_2(\beta-1)}{k_{12}\beta}\nonumber\\
k_{21}=\gamma_1-k_{12}.\label{retruc}\end{eqnarray} In other words,
all parameters are now expressed as a function of the excitation
coefficient $k_{12}$. To complete the solution we use Eqs.~17 and
19 for the single-photon rate $R$. After insertion of
Eqs.~\ref{truc} and \ref{retruc} into Eq.~17 and some lengthly
rearrangements we finally obtain
\begin{eqnarray}
k_{12}^3-\gamma_1k_{12}^2+\frac{R}{\xi}(\gamma_1+\frac{X}{k_{31}})k_{12}+\frac{RX}{\xi}=0
\end{eqnarray} with $X=(\beta-1)\gamma_1\gamma_2/\beta$. The three roots
of this cubic equation can be obtained numerically using Cardano's
method.
\section{}
We can generalize the analysis made in the previous Appendix B without
any approximation. For this we first remark  that we have the exact
relations
\begin{eqnarray}
\gamma_1\gamma_2=det[\mathbf{M}],\nonumber\\
\gamma_1+\gamma_2=-(a+d),\nonumber\\
R \cdot det[\mathbf{M}]=\xi k_{21}k_{12}k_{31}
\end{eqnarray} which can be obtained after some manipulations from
the definition of the $\textbf{M}$ matrix and of Eqs.~17,19. If we
now introduce the definitions
\begin{eqnarray}
\begin{array}{c}
A=det[\mathbf{M}]=(k_{12}+k_{21})k_{31}+(k_{12}+k_{31})k_{23},\\
B=-(a+d)-k_{31}=k_{12}+k_{21}+k_{23},\\
C=R \cdot det[\mathbf{M}]/\xi = k_{12}k_{21}\end{array}
\end{eqnarray}
we obtain after lengthly but straightforward calculations
\begin{eqnarray}
k_{12}^2-Bk_{12}+C+A-Bk_{31}=0.\label{grouik}
\end{eqnarray}
To complete the resolution of the system we use the exact
relation Eq.~\ref{co}. We finally obtain
\begin{eqnarray}
\begin{array}{c}
k_{31}=\gamma_2\gamma_1/[\gamma_1-\gamma_2)\beta+\gamma_2],\\
k_{12}=B/2-\sqrt{[B^2-4(C+A-Bk_{31})]}/2,\\
k_{21}=C/k_{12},\\
k_{23}=B-k_{12}-k_{21}. \end{array}
\end{eqnarray}
The minus sign was chosen in the second equation for $k_{12}$ (i.e.
solution of Eq.~\ref{grouik}) since a Taylor expansion at low
excitation power when $\gamma_2\ll\gamma_1$, $\beta\simeq1$ gives
for the two roots
\begin{eqnarray}
k_{12}\simeq \gamma_1/2[ 1 \pm
\sqrt{(1-\frac{4R}{\xi\gamma_1})}]\simeq
\gamma_1/2\pm\gamma_1/2\mp\frac{R}{\xi}.
\end{eqnarray} In order to have the linear regime $k_{12}\propto R/\xi$
we must therefore impose the minus sign.


\begin{thebibliography}{}
\bibitem{footnote2}
Note that for the NV\up{-} center, one should consider a four-level model by including two intermediate singlet states~\cite{Doherty1,delaney} responsible for an additional ZPL line in the IR part of the spectrum. However, because of a rapid radiative decay between these states, only the low-energy singlet plays a role in the bunching behavior.
\bibitem{footnote}
These are nominal values. Because of the uncertainty on the exact location of the NV center in the ND matrix, we expect that the exact size of the studied NDs, whether 50 or 25~nm, is not a critical parameter in the present study.
\end{thebibliography}
\end{document}